\documentclass[10pt,journal,compsoc]{IEEEtran}

\ifCLASSOPTIONcompsoc
  \usepackage[nocompress]{cite}
\else
  \usepackage{cite}
\fi

\usepackage{amsbsy}
\usepackage{amssymb}
\usepackage{graphicx}
\usepackage{color}
\usepackage{caption}
\usepackage{booktabs}
\usepackage{amsmath}
\usepackage{algorithm} 
\usepackage{algorithmic}
\usepackage{enumitem}
\usepackage[font={small}]{caption}

\usepackage{fancyhdr}
\usepackage{listings}

\setlist[itemize,1]{leftmargin=\dimexpr 25pt}

\definecolor{dkgreen}{rgb}{0,0.6,0}
\definecolor{gray}{rgb}{0.5,0.5,0.5}
\definecolor{lightgray}{rgb}{0.9,0.9,0.9}
\definecolor{mauve}{rgb}{0.58,0,0.82}
\begin{document}
%
\title{A Scenario-Based Development Framework for Autonomous Driving}

\author{Xiaoyi Li \vspace{2pt} \\ 
\IEEEcompsocitemizethanks{\IEEEcompsocthanksitem Xiaoyi (Erik) Li, 
CA, United States (e-mail: warlock.lxy@gmail.com)
}
}



\IEEEtitleabstractindextext{%
\begin{abstract}
This article summarizes the research progress of scenario-based testing and development technology for autonomous vehicles. We systematically analyzed previous research works and proposed the definition of scenario, the elements of the scenario ontology, the data source of the scenario, the processing method of the scenario data, and scenario-based V-Model. Moreover, we summarized the automated test scenario construction method by random scenario generation and dangerous scenario generation.
\end{abstract}

}

\maketitle

\IEEEdisplaynontitleabstractindextext

%
\IEEEpeerreviewmaketitle

\IEEEraisesectionheading{\section{Scenario in Autonomous Driving}\label{sec:s_scenario_in_AD}}

\IEEEPARstart{T}{he} word "scene" (Scenario) comes from the Latin Olinda, which means stage drama, and now refers to a specific situation in life. With the development of technology, the concept of scenes is gradually applied in the development and testing process of industrial production.

\subsection{Scenario Definition}\label{sec:s_scenario_definition}
Scenario-based testing was first applied to the development of software systems. ``Scenarios" were used to describe the use of the system, the requirements for use, the use environment, and the construction of more feasible systems ~\cite{carroll1995scenario, gould1987, disessa1985}. Since then, many fields have defined the term scene in their respective disciplines, such as climate change~\cite{olander2008reference}, energy industry~\cite{cayan2009climate} and so on. 

However, in the field of autonomous driving at this stage, ``scenario" has not yet been clearly defined. Since Schieben et al.~\cite{schieben2009} applied the concept of scenario to automatic driving tests, many scholars have put forward their own understanding of the term ``scenarios''. Elrofai et al.~\cite{elrofai2016scenario} defined that ``the scene is to test the continuous changes of the dynamic environment around the vehicle in a specific time range, including the behavior of the test vehicle in this environment”. Koskimies~\cite{koskimies1998automated} defined that “a scene is an informal description of a series of events when the system performs a specific task”, and an object-oriented modeling method can be used to describe the scene. RAND proposed in the autonomous driving research report that “scenarios are a combination of a series of elements used to detect and verify the behavioral capabilities of autonomous driving systems in specific driving environments”. The PEGASUS project proposes corresponding functional scenes, logical scenes, and physical scene concepts based on the differences in demand for scenes during the concept phase, system development phase, and test phase of autonomous driving product development~\cite{menzel2018scenarios}. Chinese academician Zheng Nanning of Xi’an University of Communications defines a scene as “a specific situation or scene of a traffic occasion at a specific time and in a specific space. It can be defined as a set of entities that can give a rich description of the current environment with perceptual data.”~\cite{zhengnanning2017}. Based on the above viewpoints, these scene definitions are consistent in the core elements: they all include road environment elements, other traffic participants, and vehicle driving tasks. At the same time, these elements will last for a certain period of time and have dynamic characteristics.

Therefore, the autonomous driving scenario can be understood as such: a scenario is the dynamic description of the components of the autonomous vehicle and its driving environment over a period of time. The relationship of these components is determined by the functions of the autonomous vehicle to be inspected. In short, the scene can be regarded as a combination of the driving situation and driving scene of an autonomous vehicle.

Autonomous driving scenarios are infinitely rich, extremely complex, difficult to predict, and inexhaustible. Therefore, the scenarios used for developing and testing should meet the requirements of quantifiable (the features of each element of the scenario can be quantified) and reproducible (the scenario is in the current technology The basic and test software can be reproduced) and high-fidelity (can present or reflect the real world scene to a certain extent).



\subsection{Scenario Ontology} \label{s_scenario_ontology}
Determining the ontology of the scenario element is the cornerstone of scenario-based techniques. However, there are still disputes among different researchers regarding the types and content of ontology.

\begin{figure*}[hbpt]
\centering
\includegraphics[width = \textwidth]{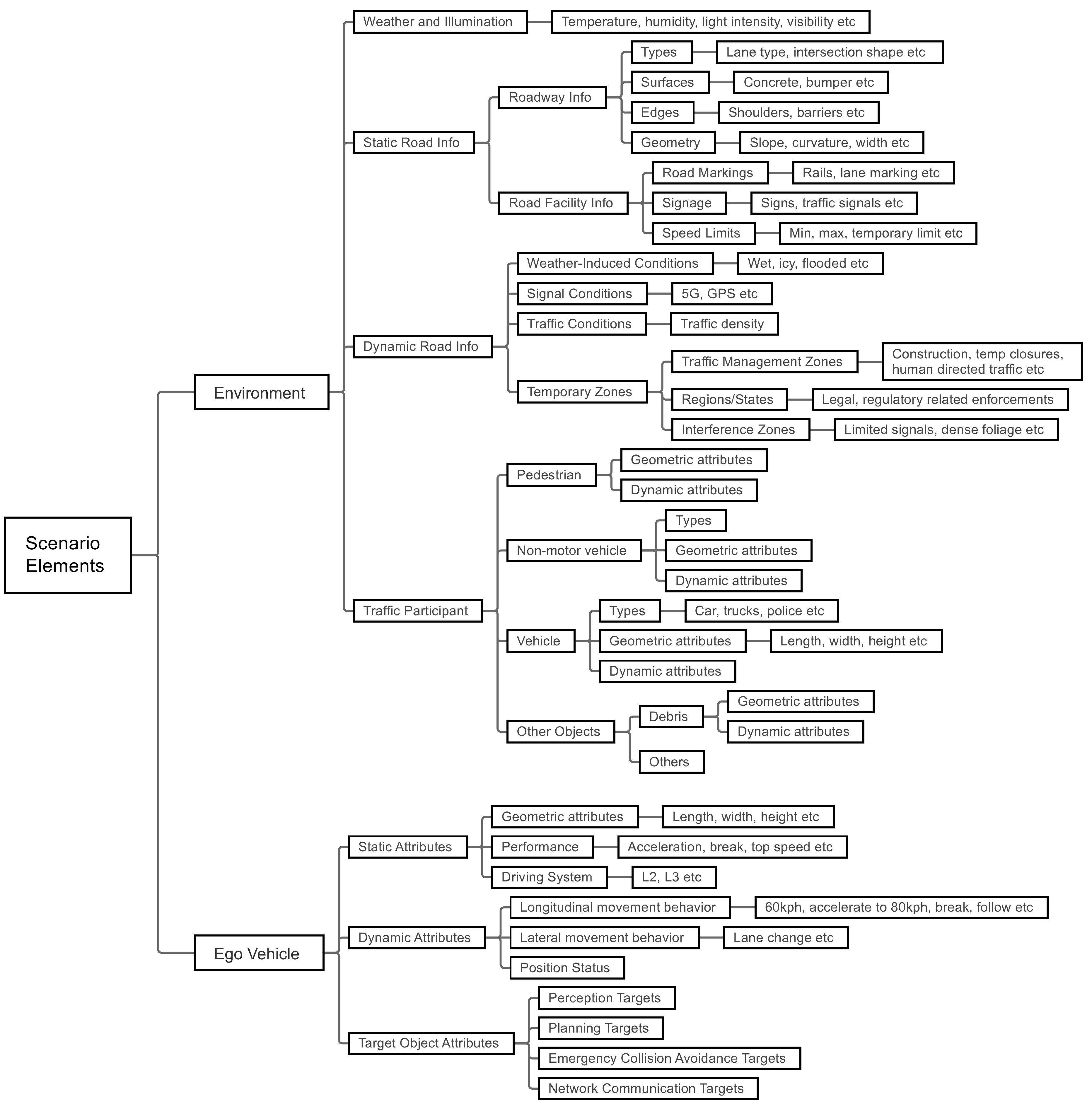}
\caption{Ontology for Scenario Elements} \label{f_ontology_elements}
\end{figure*}

Commonly used open source schemas such as OpenDrive and OpenScenario specified their road elements and traffic dynamic elements definitions in detail~\cite{10.1007/978-3-319-10584-0_11, jullien2009openscenario}. Ulbrich et al.~\cite{ulbrich2015defining} proposed that the elements of a scene should include test vehicles, traffic environment elements, driving task information, and specific driving behaviors. Autonomous driving is a part of the test scene. Geyer et al.~\cite{geyer2014} believe that the scene is the pre-defined driving environment, driving tasks, static elements and dynamic elements during the automatic driving test, and the test vehicle itself is not included in the scene. Korbinian et al.~\cite{groh2017towards} divided the scene elements into three categories: the environmental part (weather, light, wind speed, etc.), the static part (lane lines, trees, obstacles) and the dynamic part (traffic participants, pedestrians). In the latest report of RAND, the scene elements are divided into 5 layers, namely the road information layer (lane line, intersection shape, number of lanes, etc.), road infrastructure layer (traffic signs, trees, guardrails, etc.), road information layer and road The dynamic changes of the facility layer (road maintenance, tree breaking, obstacle movement, etc.), dynamic targets (pedestrians, traffic participants), environmental conditions (light, weather), test vehicles are not included. Matthaei et al.~\cite{matthaei2014map} discussed whether weather and light should be included as scene factors. Zhu et al.~\cite{ZHU2019review} categorized scenarios into test vehicles and traffic environments. Erwin et al.~\cite{de2017assessment} believe that in the early stage of system development, the scene only needs basic information about the road and other traffic participants. 

During testing, the test vehicle itself will have a significant impact on surrounding scene elements, especially other traffic participants. The interaction between the test vehicle and the surrounding driving environment forms a closed loop. At the same time, the property of the test vehicle will have a key impact on the behavioral decision-making of the automatic driving system. For example, the acceleration performance of the vehicle during overtaking plays a decisive role in the execution of the decision. Therefore, the test vehicle should be treated as a part of the scene, and the surrounding driving environment constitutes the whole scene. 

Based on this concept, we integrate the above-mentioned research and propose a scenario ontology shown in Fig \ref{f_ontology_elements}. 

In this ontology, the scenario elements have two categories: basic information of the vehicle and environment elements. Among them, the basic information of the vehicle includes three categories: basic elements of the test vehicle, target information, and driving behavior. Traffic environment elements include weather and light, Static road information, dynamic road information and traffic participant information.


\subsection{Scenario Data} \label{s_scenario_data}
It is necessary to collect a large amount of scenario data and establish a scenario library. For example, PEGASUS and KITTI in Germany, NHTSA Autonomous Driving Test Architecture Project in the United States, University of California, Berkeley BDD100K, China's "Kunlun Project", Baidu ApolloScape, etc. are all committed to providing more practical scenario data for autonomous driving research and testing~\cite{geiger2012we}.

\begin{figure}[h]
\centering
\includegraphics[width = 0.48\textwidth]{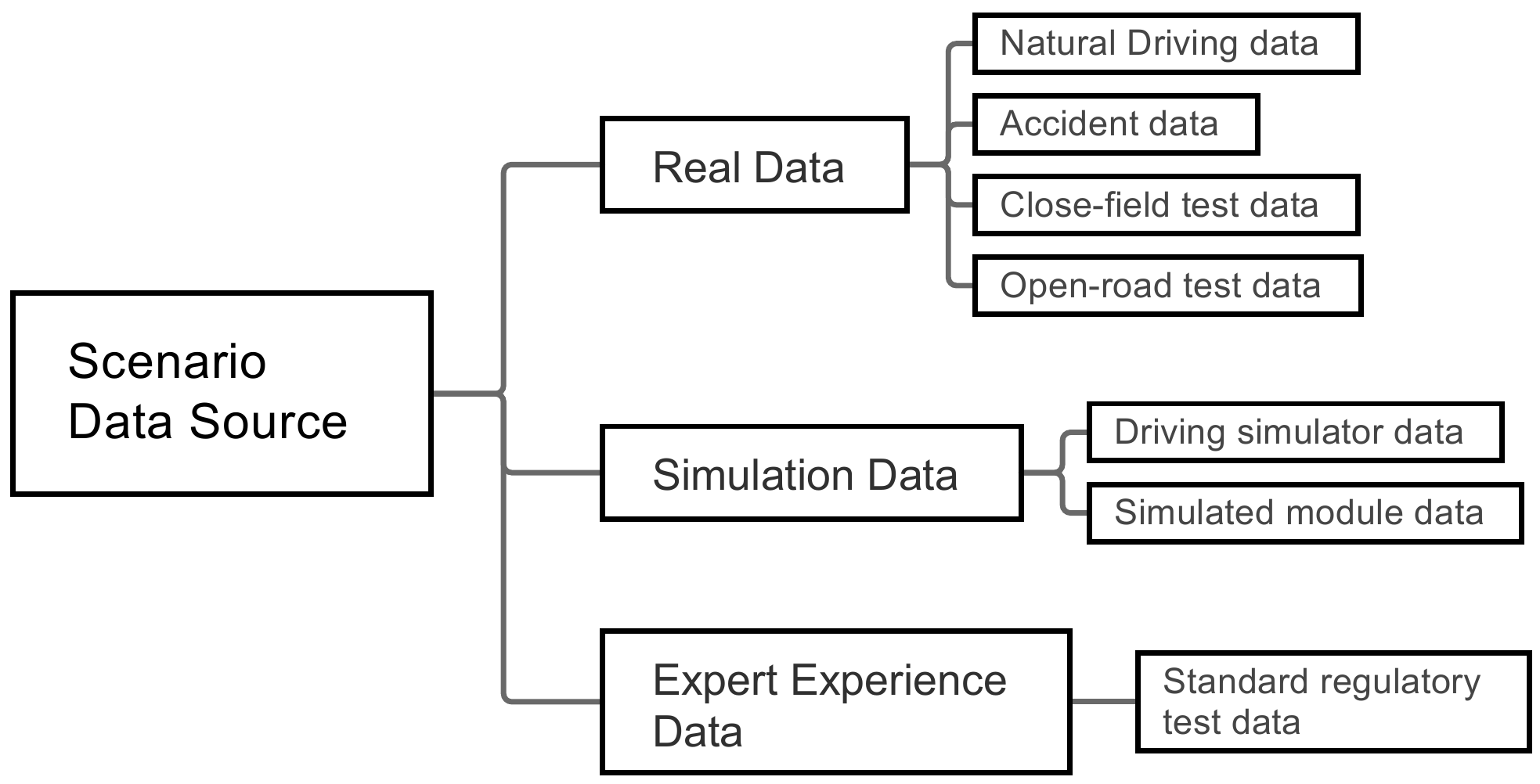}
\caption{Scenario Data Source} \label{f_scenario_data_source}
\end{figure}

The data sources mainly include three parts: real data, simulation data and expert experience data. The specific content is shown in Figure \ref{f_scenario_data_source}.

\vspace{-10pt}

\subsubsection{Real Data}
The real data sources mainly include natural driving data, accident data, close field test driving data, and open road test driving data.

The \textbf{natural driving data} is the scenario data collected during the normal driving of the vehicle by installing a multi-sensor collection platform such as radar, camera, and high-precision inertial navigation on a traditional car. Typical natural driving data collection conditions include highways, urban roads, parking lots, etc. The key to natural driving scene data collection is to ensure the time and space synchronization between sensor data. Time synchronization needs to synchronize the data collection cycles of different sensors. Currently, a unified clock source device such as GPS, COMPASS, GLONASS or GALILEO is used to achieve nanosecond synchronization between sensor data~\cite{mcinnes2009model}. For sensor data of different frequencies, median sampling, spline difference sampling and other methods can be used to achieve time synchronization~\cite{sivrikaya2004time}.

The \textbf{accident data} is the scenario data refined using the existing big data of road traffic accidents. At this stage, many countries and organizations have established traffic accident databases, such as China’s CIDAS database, Germany’s GIDAS database, US NHTSA’s GES database, and EU’s ASSESS database, etc. Automated driving tests can make full use of the data resources provided by these databases to construct test scenarios based on traffic accidents and illegal scenarios.

\subsubsection{Simulation Data}
Simulation data refers to the test data obtained by virtual operation of the autonomous vehicle in a simulation environment. The simulation environment can be generated through real scene import or vehicle driving environment modeling. Vehicle driving environment modeling mainly includes road scene modeling, traffic environment modeling, weather modeling and electromagnetic environment modeling. The key to traffic environment modeling is to generate correct traffic flow information and the behavior of surrounding traffic vehicles. At present, cellular automats are mostly used. Meteorological modeling and electromagnetic environment modeling aim to restore the weather conditions and electromagnetic interference in the real environment, such as simulating light intensity, humidity, temperature, shadow effects of electromagnetic signals, Doppler frequency shift, etc. 

\subsubsection{Expert Experience Data}
Expert experience data refers to the scene element information obtained through the experience and knowledge of the previous tests. At present, there are more than 80 types of autonomous driving test laws and regulations in countries around the world. Taking the Autonomous Emergency Braking (AEB) function as an example, Euro-NCAP divides the AEB function test into three types: AEB-City, AEB Inter Urban and AEB Pedestrian~\cite{park2016design}, each test type has its corresponding test scenario.

\subsection{Scenario Data Processing} \label{s_scenario_data_processing}
The key to scene data processing is the deconstruction and reconstruction of scene elements.

The German PEGASUS project proposes 7 steps for scene data processing~\cite{Ptz2017SystemVO}: Generate a general environment description, check the data format, generate additional information, analyze the degree of correlation between the scenes, analyze the possibility of scene occurrence, cluster logical scene data and calculate the frequency distribution, and generate specific test scenes based on the generated logical scenes. Baidu proposed a three-step method of scene clustering including scene classification rule definition, scene labeling (element decomposition, quantification), and label clustering. 

\begin{figure}
\centering
\includegraphics[width = 0.46\textwidth]{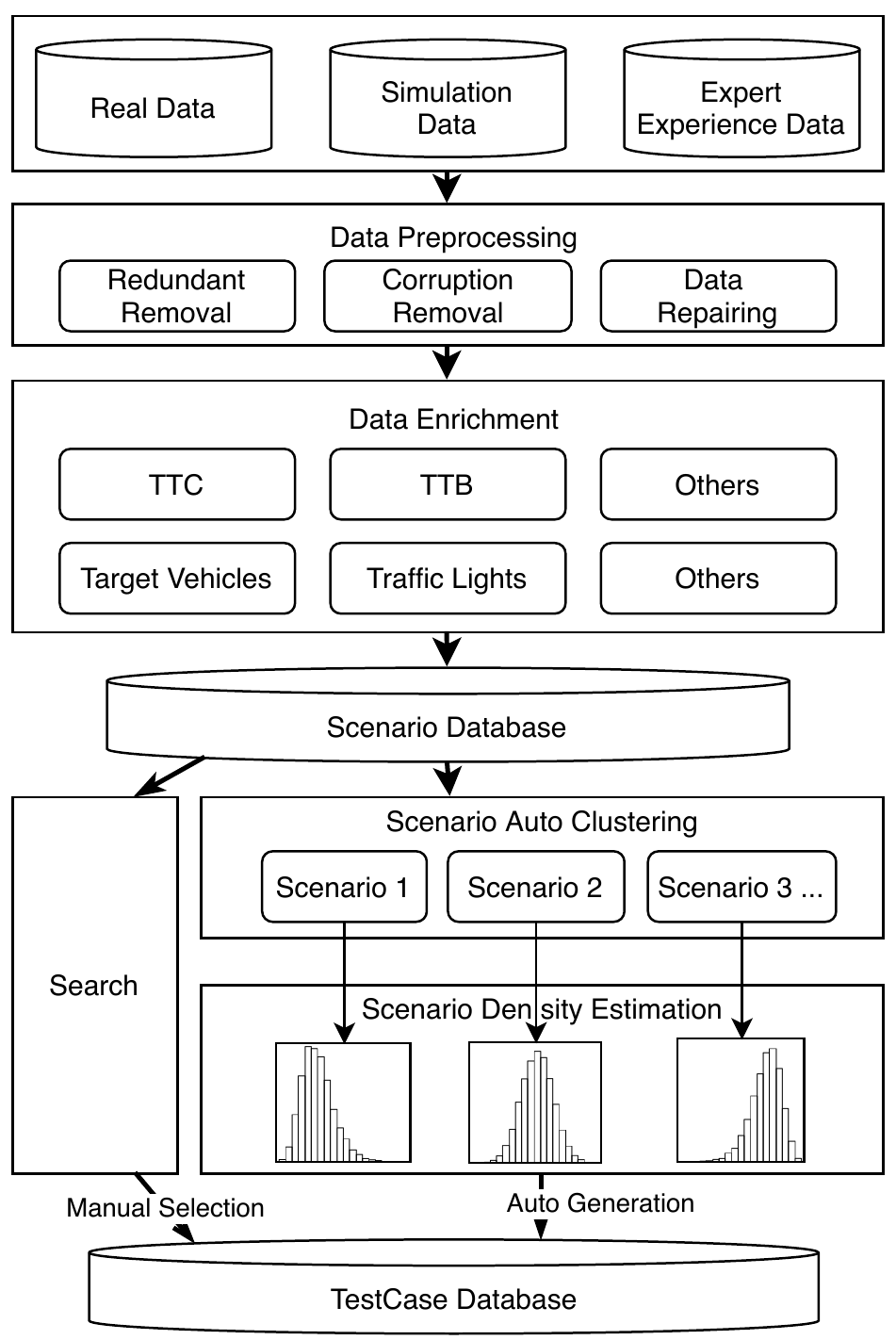}
\caption{Scenario Data Processing Flow} \label{f_data_pipeline}
\end{figure}

According to the existing typical scene data processing methods, this article summarizes and proposes the scene data processing flow shown in Figure \ref{f_data_pipeline}.

\subsubsection{Data Preprocessing}
Sensor data from different channels are multimodal. There are also a lot of invalid data and misaligned data in the original data. Therefore, the sensor data cleaning becomes the prerequisite to construct the scenario library. 
 
Cleaning the collected scene data mainly includes removing redundancy, deleting missing data, data repairing, etc. The data repairing can be done manually by completing key information or repairing according to the statistical value of the data. In the data cleaning process, it should meet the following requirements: maintain the data integrity; formulate user customized cleaning rules; minimize cleaning cost~\cite{fan2008conditional, fan2012towards}. Taking data restoration as an example, the cleaning cost is measured by reconstruction error $\mathcal{J}$, which is defined as:

\begin{equation}  \label{e_cleaning_cost}
	\mathcal{J} = \frac{1}{n}\sum_{i=1}^{n}\mathcal{D}(g(X_i),X_i),
\end{equation}
where the $g(x)$ means any reconstruction methods, $\mathcal{D}$ is the distance function where \textit{Damerau-Levenshtein} distance is usually used. The cleaned data is then organized to form a usable scene dataset.

\subsubsection{Data Enrichment}
Cleaned data will be enriched internally and externally. Internally, additional information can be derived from data directly, including the calculation for time-to-collision (TTC), time headway, time-to-brake (TTB) and etc~\cite{hallerbach2018simulation}. Externally, key information in the data is annotated by external annotators. Annotators can be human-based or algorithm-based (a.k.a Auto Annotation). Commonly used algorithms include supervised and semi-supervised methods~\cite{belkin2006manifold, wang2018extracting, yang2003automatically, murthy2015automatic}. 

\subsubsection{Scenario Clustering}
Annotated scenarios is clustered based on Ontology. The scenes that meet the classification criteria are clustered into corresponding scene elements, and the parameter space of scene elements is clarified. Commonly used clustering algorithms mainly include K-Means clustering, hierarchical clustering, Gaussian Mixture model, Deep learning based clustering such as T-SNE~\cite{hinton2003stochastic}.

\subsubsection{Scenario Density Estimation}
Based on the above clustered scenarios, the kernel density functions of the ontology scenarios are calculated to facilitate the random generation of specific scenarios in section \ref{s_random_scenario_gen}. Suppose $x_1, x_2,..., x_n$ are $n$ scenarios with independent and identical distribution. Let its probability density function be $f$, the kernel density function estimator is defined as:

\begin{equation}  \label{e_kernal_density}
	f_h(x) = \frac{1}{n}\sum_{i=1}^{n}\mathcal{K}_h(x - x_i),
\end{equation}
where 
\begin{equation}  \label{e_k_density}
	\mathcal{K}_h(x) = \frac{1}{h}\mathcal{K}(\frac{x}{h}).
\end{equation}

In this estimator, $\mathcal{K}$ is the kernel function, non-negative and the integral value is 1. $h$ is the smoothing factor, which is determined by the square error of the average integral; $\mathcal{K}_h$ is the smoothed kernel function. With these density functions, test-cases can be manually picked or randomly generated according to $\mathcal{K}$ of specific scenarios.

\section{Scenario-based V-Model} \label{s_scenario_based_v_model}
\begin{figure}[h]
\centering
\includegraphics[width = 0.48\textwidth]{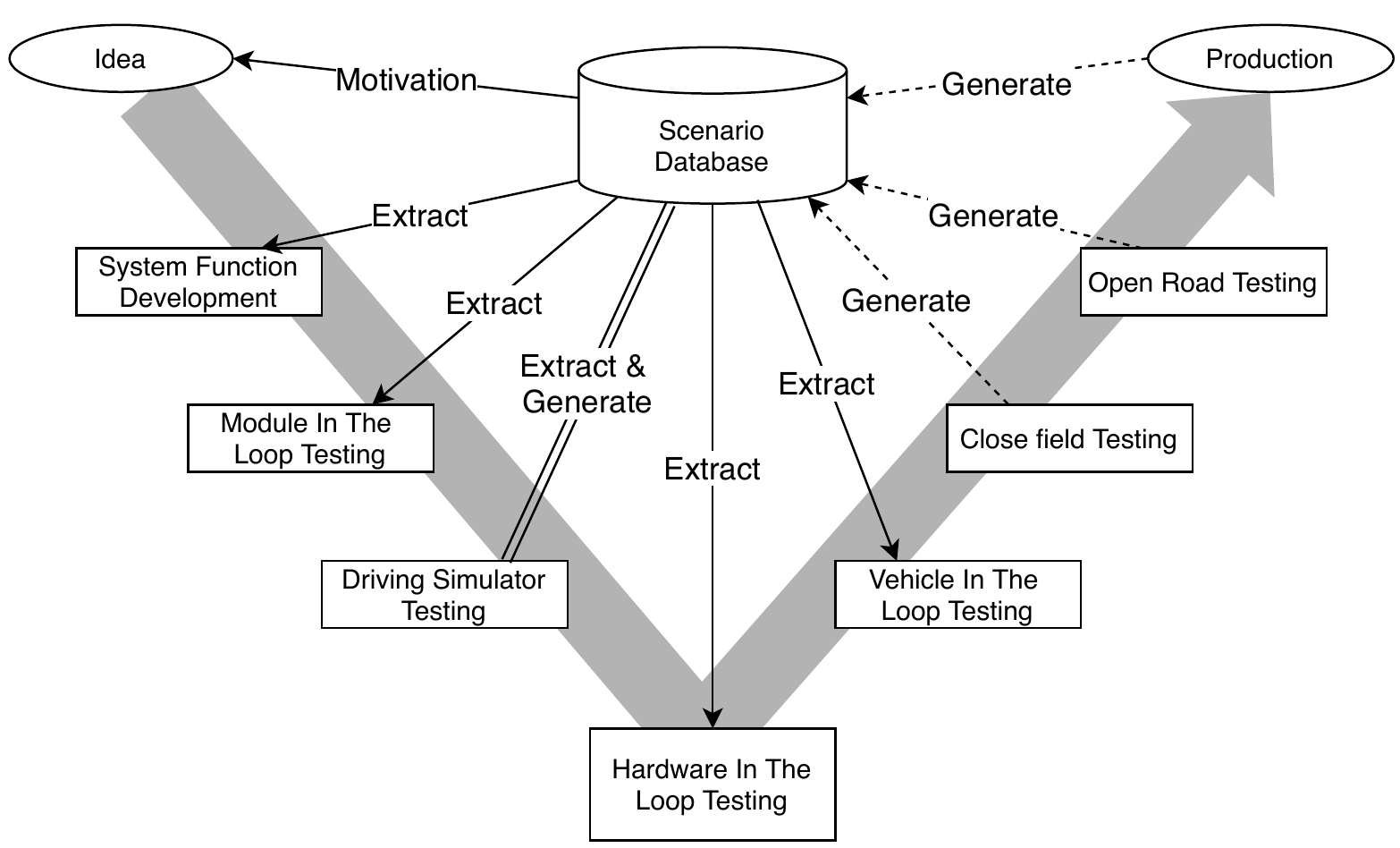}
\caption{Scenario-based V-Model. The scenario database is embedded in all stages of the development, where the scenario extraction is mostly achieved by search interfaces.} \label{f_v_model}
\end{figure}

With the level of autonomous driving increases, the test scenarios become infinitely rich, extremely complex, unpredictable, and inexhaustible. Covering all situations in road testing is no longer possible. A scenario-based V-Model testing framework is shown in Figure \ref{f_v_model}. It includes virtual testing, such as software-in-the-loop testing (SIL), hardware-in-the-loop testing (HIL), and real road testing, such as close field testing and open road testing~\cite{lattarulo2017complete, gonzalez2015review, bagschik2018ontology}. 

Car companies and research institutions are gradually pursuing scenario-based direction due to the abundant scenarios, fast calculation speed, high test efficiency, low resource consumption, good repeatability, and easy embedding in all aspects of vehicle development. The scenario property among virtual testing, close-field testing and open-road testing are summarized in table~\ref{t_scene_property} below.

\vspace{10pt}
\begin{table}[h]
\begin{center}
\begin{tabular}{|l|l|l|}
\hline
 &
  \multicolumn{1}{c|}{\textbf{Number of Scenarios}} &
  \multicolumn{1}{c|}{\textbf{How are they used}} \\ \hline
\textbf{\begin{tabular}[c]{@{}l@{}}Virtual \\ Testing\end{tabular}} &
  \begin{tabular}[c]{@{}l@{}}High. Any number of \\test scenarios can be \\ generated in the given \\logical scenario \\ parameter space\end{tabular} &
  \begin{tabular}[c]{@{}l@{}}\\ Embedded in all aspects of \\ system development, conduct \\ massive scene tests to verify \\ the boundaries of autonomous \\ driving functions \\\\ \end{tabular} \\ \hline
\textbf{\begin{tabular}[c]{@{}l@{}}Close\\ Field \\ Testing\end{tabular}} &
  \begin{tabular}[c]{@{}l@{}}Low. Due to the \\ limitation of the test \\ field.\end{tabular} &
  \begin{tabular}[c]{@{}l@{}}\\ Validate key scenes, and \\ build new scene types that are \\ not encountered or have low \\ probability \\\\ \end{tabular} \\ \hline
\textbf{\begin{tabular}[c]{@{}l@{}}Open\\ Road \\ Testing\end{tabular}} &
  \begin{tabular}[c]{@{}l@{}}High. Can encounter \\as many required test \\ scenarios as possible\end{tabular} &
  \begin{tabular}[c]{@{}l@{}}\\ Clarify the statistical property \\ of related events, verify the \\ system boundaries in actual \\ situations, detect the interaction \\ between autonomous vehicles \\ and traditional vehicles, and \\ discover new scenarios \\ that have not been considered \\\\ \end{tabular} \\ \hline
\end{tabular}
\caption {Property of scenario in different testing stages} \label{t_scene_property}
\end{center}
\end{table}
\vspace{-10pt}

\section{Automatic Scenario Generation}
As shown in Figure \ref{f_data_pipeline}, when we don't have enough scenarios to do SIL testing, we have to generate scenarios by human or machine. Human expert can generate very customized scenarios for testing. However, the cons are obvious -- expensive and unscalable. The goal of this section is to automatically generate a large number of test scenarios in a short time according to test requirements. The generation methods mostly fall into two categories: random scenario generation and dangerous scenario generation.

\subsection{Random Scenario Generation} \label{s_random_scenario_gen}
Based on the probability density $\mathcal{K}$ of various scenes in Eq.\ref{e_kernal_density}, specific scenes can be randomly generated in the virtual environment. The generation methods mainly lies in three categories. 1) Random sampling represented by Monte Carlo sampling and fast random search tree. 2) Importance based sampling such as importance level analysis of scene elements. 3) Machine learning based methods.

\subsubsection{Random Sampling}
Yang et al.~\cite{yang2010development} and Lee~\cite{lee2004longitudinal} extracted data fragments from road collision pre-warning and adaptive cruise field tests, then used Monte Carlo simulation to generate the test scenario for `active braking'. Olivaves et al.~\cite{olivares2016virtual} used Markov chain Monte Carlo methods to reconstruct road information by analyzing road map data. Fellner et al.~\cite{fellner2019model} applied the Rapidly-exploring Random Tree (RRT) method in path planning to scene generation, and the generated test cases can consider more than 2300 scene elements. Li et al.~\cite{li2019spatiotemporal} proposed a common model construction method based on road image sequence, which uses Super-pixel Markov random field algorithm to monitor the road area and realize the random modeling of the road scenario. Elias et al.~\cite{rocklage2017automated} proposed a scene generation method based on the backtracking algorithm, which can randomly generates dynamic and static scene elements.

\subsubsection{Importance Based Sampling}
Importance based sampling~\cite{xia2017automatic} usually contains three major steps. First, it needs to analyze the scene elements, clarify the scene elements, and discretize the continuous scene elements. Then determine the importance score of each scene element through information entropy and level analysis. Next, the importance score of different elements is flattened, and the relative importance parameters of each scene element are obtained. Finally, testcases are generated through the combined test scenarios.

\subsubsection{Machine Learning Based Sampling}
Schillinng et al.~\cite{schilling2016validation} approached the problem by changing the nature of scene elements, such as white balance, light changes, motion blur, etc. Alexander et al.~\cite{koenig2017bridging} infer the behavior information of surrounding traffic participants based on the collected data, and use neural networks to learn the behavior information of surrounding vehicles to generate dynamic scenes. Li et al.~\cite{huang2018study} divided the driving position around the car into 8 areas, then generate scenarios through the arrangement and combination of the relative position and speed of the vehicle and the surrounding traffic vehicles. Vishnukumar et al.~\cite{Vishnukumar2017MachineLA} proposed to apply the deep learning method to the test verification process. After the initial necessary test scenarios are given, random test scenarios are automatically generated through learning algorithms.

\subsection{Dangerous Scenario Generation}
Compared with building real test scenarios in the real world, generating test cases in a virtual environment can greatly reduce time and resource consumption. However, due to the low probability of accidents under natural circumstances, the method of using random generation may still face a large number of calculation difficulties. Putting more weight on dangerous scenes generation can alleviate this problem.

First of all, it is necessary to define and classify dangerous scenes. Many projects have conducted research on car dangerous scenes. SeMiFOT divides the risk of driving into 4 levels~\cite{ahlstrom2011processing}. The United States NHTSA classifies collisions into 37 categories~\cite{najm2007pre}. Aparicio et al.~\cite{aparicio2013status} summarized the types of conflicts between cars and cars, cars and pedestrians. Winkle et al.~\cite{winkle2018area} analyzed accident data in which the line of sight was blocked in different weather conditions from 2004 to 2014, and analyzed the severity of the accident.

The definition of the above-mentioned dangerous scenes is narrow where most of them only analyze the types of their dangers without defining specific parameters of the scene elements. Tang et al.~\cite{tang2011development} define each attribute parameter of the accident scene, and propose a method for drawing urban traffic accidents. Sven et al.~\cite{hallerbach2018simulation} used specific parameters such as TTB, expected braking deceleration, TTC, traffic flow, speed fluctuation, average speed, acceleration change and other specific parameters to find the dangerous scenes from the massive car driving data. Elrofai et al.~\cite{elrofai2016scenario} judged whether there is lane changing behavior by detecting the speed and yaw rate of the vehicle during driving. When the continuous yaw rate exceeds the threshold for a period of time, it is judged as a valuable steering event. Huang et al.~\cite{huang2017evaluation} proposed a method to accelerate the generation of dangerous scenes based on importance sampling based on the defined dangerous scenes. The core idea is to introduce a new probability density function $f^*(x)$ to increase the probability of producing dangerous scenes, thereby reducing the number of tests. When using the randomly scene generation method, the probability density function of the dangerous scene is $f(x)$, and the minimum number of tests $n$ is

\begin{equation}  \label{e_min_num_test}
	n = \frac{z(1-\gamma)}{\gamma},
\end{equation}
where $\gamma$ is the probability of a dangerous scenario, $z$ is related to the inverse cumulative distribution function of $\mathcal{N}(0,1)$. 

When importance sampling is used to generate dangerous scenes, the probability density function of the dangerous scenes is $f^*(x)$, and the minimum number of tests is
\begin{equation}  \label{e_new_min_num_test}
	n = z \begin{Bmatrix}
\frac{E_{f^*}[I^2(x)\cdot L^2(x)]}{\gamma^2 - 1} - 1
\end{Bmatrix},
\end{equation}
where $I(x) \in [1, 0]$ is the index function of dangerous event $\varepsilon$ and $L(x) = \frac{f(x)}{f^*(x)}$ is likelihood ratio for using importance sampling. $E_{f^*}[I^2(x)\cdot L^2(x)]$ is the probability of occurrence of the dangerous scene after changing the probability density function to $f^*(x)$. 

Through the verification of the method for typical scenes such as cut in and AEB, it is proved that the test speed is 7,000 times faster than Monte Carlo test simulation.

\subsection{Technical Challenges}
There are three technical challenges for auto test scenario generation: authenticity, granularity, and measurement.

\subsubsection{Authenticity}
In order to ensure the authenticity of the scene during the virtual test, the reference measurement system (RMS) should be established during the virtual scene test~\cite{leitner2018challenges}. RMS is mainly used to compare the difference between the generated virtual test scene and the real world. Its accuracy needs to be higher than that of sensors on autonomous vehicles. If the roughness of the scene elements detected by the RMS system is less than a certain a threshold value, it can prove that the generated virtual test environment can be used to test the automatic driving function. Taking the lane keeping function as an example, the necessary environmental element information includes road shape, lane line position, lane line shape, and light conditions. At this point, the main component of the RMS is the image acquisition device, which has better performance in terms of resolution and sensitivity than the sensors used in autonomous vehicles. The RMS image acquisition device is the placed on the HIL test bench built above for detection. If the detected road color features, lane line gray value, lane line edge shape and other characteristics are similar to the real world, it proves that the fidelity of the generated virtual scene meets the requirements.

\subsubsection{Granularity}
The granularity of scene elements needs to be adapted according to technological development. Taking the size of raindrop particles as an example. The size of raindrops will cause greater interference to radar echo. The smaller the raindrops, the weaker the reflection of microwaves. For radar, when the diameter of raindrops is less than a certain threshold, the detection results of the radar will almost remain unchanged for the decision-making results of the entire autopilot system. At this time, blindly pursuing the reality of simulation, such as reducing the particle size of raindrops, will increase The consumption of large calculations puts a great burden on the computation of the simulation system. Therefore, the authenticity of the simulation environment needs to consider the technical level of the sensors currently used and the computing power.

\subsubsection{Measurement}
Collision is often used as the measurement for the virtual test. In order to increase the virtual test coverage, Tong et al.~\cite{tong2017simu} proposed a way of specifying key performance indicators (Key Performance Indicator, KPI) to describe the performance of autonomous vehicles. Taking the adaptive cruise system as an example, the KPI parameters describing the adaptive cruise performance in the virtual test include: safety (the ability to avoid collisions), comfort (vehicle acceleration and deceleration), naturalness (the similarity of human driving), economy (fuel consumption), according to different automatic driving functions, different KPIs can be set for evaluation. Some scholars have also proposed the use of the Turing test as measurement. Li et al.~\cite{li2019parallel} proposed a driver-in-the-loop parallel intelligent test model, which uses the principle of Turing test to test the understanding of the elements and driving decision-making capabilities of autonomous vehicles in complex scenarios.

\section*{Disclaimers}
Draft for open concept instruction. Algorithms are partial and figures are subject to change.

\ifCLASSOPTIONcaptionsoff
  \newpage
\fi


\bibliographystyle{abbrv}
\bibliography{all_citation}

\newpage

\end{document}